\documentclass[%
 reprint,
nofootinbib,
 amsmath,amssymb,
 aps,
]{revtex4-2}

\usepackage{aasmacros}
\usepackage{graphicx}% Include figure files
\usepackage{dcolumn}% Align table columns on decimal point
\usepackage{bm}% bold math
\usepackage{ulem}
\usepackage{xcolor}
\usepackage{hyperref}

\begin{document}

\preprint{APS/123-QED}

\title{Computation of stochastic background from extreme mass ratio inspiral populations for LISA}

\author{Federico Pozzoli}
\affiliation{Dipartimento di Scienza e Alta Tecnologia, Universit\'{a} degli Studi dell’Insubria, I-22100 Como, Italy}

\author{Stanislav Babak}%
\affiliation{%
APC, AstroParticule et Cosmologie, CNRS, Universitè de Paris, F-75013 Paris, France}%

\author{Alberto Sesana}
\affiliation{Dipartimento di Fisica “G. Occhialini”, Universit\'{a} degli Studi di Milano-Bicocca, 20126 Milano, Italy
}%

\author{Matteo Bonetti}
\affiliation{Dipartimento di Fisica “G. Occhialini”, Universit\'{a} degli Studi di Milano-Bicocca, 20126 Milano, Italy
}%

\author{Nikolaos Karnesis}
\affiliation{
Department of Physics, Aristotle University of Thessaloniki, Thessaloniki 54124, Greece}%

\begin{abstract}
Extreme mass ratio inspirals (EMRIs) are among the primary targets for the Laser Interferometer Space Antenna (LISA). The extreme mass ratios of these systems result in relatively weak GW signals, that can be individually resolved only for cosmologically nearby sources (up to $z\approx2$). The incoherent piling up of the signal emitted by unresolved EMRIs generate a confusion noise, that can be formally treated as a stochastic GW background (GWB). In this paper, we estimate the level of this background considering a collection of astrophysically motivated EMRI models, spanning the range of uncertainties affecting EMRI formation. To this end, we employed the innovative \textit{Augmented Analytic Kludge} waveforms and used the full LISA response function. For each model, we compute the GWB SNR and the number of resolvable sources. Compared to simplified computations of the EMRI signals from the literature, we find that for a given model the GWB SNR is lower by a factor of $\approx 2$ whereas the number of resolvable sources drops by a factor 3-to-5. Nonetheless, the vast majority of the models result in potentially detectable GWB which can also significantly contribute to the overall LISA noise budget in the 1-10 mHz frequency range.

\end{abstract}
\maketitle

\section{Introduction}
Galactic nuclei are among the densest structures in the Universe, and it has been showed that they host a massive black hole (MBH) in their center. At such high densities, generally exceeding $10^6 \rm M_{\odot} pc^{-3}$, the high rate of strong and weak gravitational encounters among stars and compact objects (COs) efficiently redistribute their energy and angular momentum, occasionally resulting in close encounters with the central MBH. From this stems a wide variety of spectacular phenomena driven by extreme dynamics such as stellar tidal disruptions \cite{Rees1988}, hypervelocity stars ejection \cite{Hills1988}, relativistic captures of COs \cite{Amaro-Seoane2018} and quasi-periodic eruptions \cite{King2022}. 

Specifically, when scattered on very low angular momentum orbits, COs decouple from the influence of the stellar environment and, together with the central MBH, evolve as a relativistic binary approximately in isolation. The energy of the binary is gradually released through the emission of gravitational waves (GWs), causing the inspirals of the CO onto the MBH. Considering the enormous difference between the two masses of the binary (typically $1-50 \rm M_{\odot}$ for the CO and $10^5-10^9 \rm M_{\odot}$  for the MBH), such events are called Extreme Mass Ratio Inspirals (EMRIs), with mass ratio in the range $q=\mu/M=10^{-9}-10^{-4}$, being $M$ the mass of the MBH and $\mu$ the mass of the CO. 
Several channels of EMRI formation have been proposed, which modify the process described above either by adding further physical effects such as resonant relaxation and BH-BH scattering events \cite{Amaro-Seoane2018,AmaroSeoane2007}, or by invoking different formation processes,
like binary tidal separation \cite{Miller:2005}, captures of cores of giants \cite{Distefano:2001}, massive star capture or production in accretion discs \cite{Levin:2007}.

The GW emitted by EMRIs is a primary target for the forthcoming space mission LISA (Laser Interferometer Space Antenna) \cite{LISA2017}. The observatory consists of three spacecrafts forming an equilateral triangle with $\sim 2.5 \times 10^6 \rm km$ arm-length. Lasers will be sent both ways between each pair of spacecrafts, and it will measure the phase changes between the transmitted and the received beam, induced by the deformation of the space-time due to the passing GW. The LISA sensitivity will span the $0.1-100\rm mHz$ frequency range, enabling the detection of GWs produced by different kind of events such as massive black hole binary (MBHB) coalescences \cite{Colpi2019}, galactic compact binaries (GCB) \cite{Nelemans2001}, inspiralling stellar origin black hole (SOBH) binaries \cite{2016PhRvL.116w1102S} and EMRIs.

Due to their tiny mass ratio, EMRIs evolve slowly, completing $\sim 10^4-10^6$ cycles in LISA’s sensitive frequency range before eventually plunging onto the central MBH \cite{Hinderer2008,Peters1964}. A large number of cycles allow measuring the parameters of the binary with exquisitely high precision. Therefore, EMRIs are ideal sources to map MBH spacetime  \cite{Ryan1995,Barack2007}, perform tests of General Relativity \cite{Gair2013}, and possibly detect the presence of gas around the central MBH \cite{Barausse2007,Barausse2015}. Measuring the properties of a population of EMRI signals could additionally provide information on the mass distribution of MBHs \cite{Gair2010} and their host stellar environment \cite{AmaroSeoane2007}.
Concerning waveform modelling, the large number of orbital cycles requires an extremely faithful waveform, since even a tiny dephasing over one cycle can jeopardize the correct signal recovery through matched filtering \cite{Babak2010}. Additionally, the vast majority of EMRIs are not expected to be individually detectable with LISA because they are too far away or too far from the final plunge onto the MBH. Therefore, thousands of EMRIs will be present in the LISA data without reaching the 'individual detection threshold' and their GW emission can pile up incoherently, forming a confusion noise \cite{Barack2004-1,Bonetti2020}. In the worst case scenario, this stochastic gravitational wave background (GWB) could even overcome the level of the anticipated instrumental noise, possibly jeopardizing the detection of other sources. This is, for example, the case with the collective signal from unresolvable GCBs, which constitutes the primary  astrophysical limitation for the LISA mission in the frequency range $[0.2,3]\,\rm mHz$. The GCB confusion noise is expected to be highly anisotropic, unlike the EMRIs one, and its amplitude in the LISA detector fluctuates due to the satellite constellation motion. This modulation makes this signal at least partially subtractable from the LISA error budget when searching for an underlying stochastic background. 

Despite the characterization of GWB from EMRI is essential for the fulfillment of the LISA goals, it has so far raised little attention in the vast literature involving LISA. The first estimation has been done by the pioneering work of \cite{Barack2004}, which used  basic piece-wise approximations for the inclination and eccentricity averaged GW signal from unresolved EMRIs  and considered early estimates of the EMRI rates, in terms of a redshift independent MBH mass function.  Subsequently, \cite{Bonetti2020} improved the estimate by considering the EMRI populations of \cite{Babak2017}, but still employing a straightforward waveform for the GW signal, namely a simplified version of the Analytic Kludge (AK) used by \cite{Barack2004}. 

In this study, we want to contribute by adding further elements to make the estimate more credible. To provide a realistic estimate, we couple the astrophysical motivated model developed in \cite{Babak2017} to the Augmented Analytic Kludge \cite{Chua:2020} waveform, calibrated against Numerical Kludge model \cite{Babak2008}. Most importantly, we perform the first realistic injection of the collective distribution of EMRI signals in LISA by directly computing the GW response of the instrument in the Time Delay Interferometry (TDI) channels. We use the recursive algorithm developed in \cite{Karnesis:2020} to subtract resolvable sources and estimate the residual GWB level. Our approach not only ensures a better GWB characterization, but it also contributes to the development of tools for realistic signals injection into the LISA data stream, which is critical for future investigations and for the development of appropriate analysis pipelines in the context of the LISA Consortium.

The plan of the paper is as follows. In Section \ref{II}, we provide all the technical aspects concerning the LISA sensitivity curve and the TDI technique. In Section \ref{III}, we briefly review Kludge waveforms used to model the GW signal from EMRIs, highlighting their differences and justifying the use of AAK. Section \ref{IV} describes the astrophysical models chosen to construct the EMRI populations and defines the corresponding catalogues. Finally, the two last Sections \ref{V} and \ref{VI} are dedicated respectively to presenting the results and to drawing our final conclusions.  Throughout the paper, we adopt geometrical units $G=c=1$.

\section{LISA}\label{II}

\subsection{Sensitivity}
The definition of the sensitivity is closely related to the signal-to-noise ratio SNR which for a deterministic source can be defined as:
\begin{equation}
    {\rm SNR}^2 = 4Re \left( \int^{\infty}_0 df\frac{\tilde{h}(f)\tilde{h}^*(f)}{S_n(f)}\right),
\end{equation}
where we have introduced the one-sided noise power spectral density (PSD) $S_n(f)$ and the signal $\tilde{h}(f)$ in frequency domain.  The detector noise is described as a stochastic variable $n(t)$, and the noise PSD is defined as 
\begin{equation}
    E[\tilde{n}(f)\tilde{n}(f')] = \frac{1}{2}\delta(f-f')S_n(f)
    \label{eq2}
\end{equation}
where E[\dots]  is the expectation value, $\tilde{n}(f)$ is the Fourier transform of $n(t)$, and $\delta$ is the Dirac delta distribution.

In the analysis, we adopt the noise model of ESA’s science requirement document \cite{LISA2018}. That noise model has been approximated into two effective
functions. The high frequencies components are represented by the readout noise $S_{\rm OMS}$, whereas the low-frequency components by the single mass-acceleration noise $S_{\rm acc}$. 
Each LISA spacecraft hosts a test mass designed to follow geodesic motion only. However,  LISA Pathfinder showed that they are not perfectly free-falling. External spurious forces push them out of their geodesics, resulting in non-inertial residual accelerations, which are collected in $S_{\rm acc}$. Instead, $S_{\rm OMS}$ collects all the optical metrology system noise entering via power measurements at the photodetectors. $S_{\rm OMS}$ and $S_{\rm acc}$ are respectively given by:
\begin{equation}
S_{\rm OMS}(f) = (1.5\cdot 10^{-11})^2 \left[ 1 + \left(\frac{2\rm mHz}{f}\right)^4\right]\left( \frac{2\pi f}{c}\right)^2 \rm Hz^{-1}
\end{equation}
and 
\begin{equation}
\begin{split}
S_{\rm acc}(f) =& (3\cdot 10^{-15} m \hspace{1mm}s^{-2})^2 \left[ 1 + \left(\frac{0.4\rm mHz}{f} \right)^2\right]\\
&\left[ 1 + \left(\frac{f}{8\rm mHz} \right)^4\right]\left( \frac{1}{2\pi f} \right)^4 \rm Hz^{-1}.
\end{split}    
\end{equation}
Combining these two equations, the total noise in a Michelson-style LISA data
channel is:
\begin{equation}
    S_n (f) = \frac{S_{\rm OMS}}{L^2} + 2\left[ 1 + \cos^2 \left(\frac{f2 \pi L}{c} \right)\right]\frac{S_{\rm acc}}{L^2},
\end{equation}
where $L$ is the length of the arm equal to $2.5\times10^6$km. We highlight that \cite{Bonetti2020}, besides the sky-averaged sensitivity for LISA, took also into account the confusion noise generated by unresolved GCB (mostly white dwarf, WD, binaries). That component has been added according to the fitted formula obtained in \cite{Karnesis:2020}. Generally, the GCB population produces a stochastic GWB that effectively degrades the instrumental sensitivity at frequencies below $1 \,\rm mHz$. Here, we ignore the CGB GWB to  produce robust estimates that depend on the LISA noise only. Although including the CGB GWB might affect the EMRI GWB observability, the main contribution to the SNR of the latter comes from $f>2\,\rm mHz$, where the former should have a minor impact.

\subsection{Time Delay Interferometry} \label{sec2.b}
The main source of noise for LISA will be laser frequency noise. It’s due to the frequency instability of the onboard lasers. Ideally, a  laser operates at a single frequency with zero linewidth. In the real world, however, a laser has a finite linewidth because of phase fluctuations, which cause instantaneous frequency shifts away from the central frequency.
Equal-arm interferometer detectors can observe GW by cancelling the laser frequency fluctuations affecting the light injected into their arms. This is done by comparing phases of split beams propagated along the equal arms of the detector. The laser frequency fluctuations affecting the two beams undergo the same delay and cancel out at the photodetector. However, for a not-equal arms
interferometer (as LISA) the exact cancellation of the laser frequency fluctuations does not take place at the photodetector. For this reason, a post-process technique has been developed. The algorithm is called Time Delay Interferometry, and it was first proposed by \cite{Tinto2004}.
\begin{figure}
    \centering
    \includegraphics[width=\columnwidth]{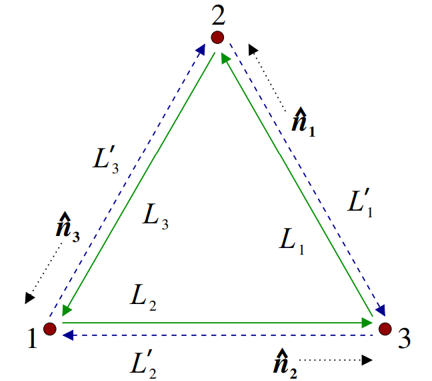}
    \caption{Schematic LISA configuration. The image has been taken from \cite{Tinto2021}. }
    \label{fig1}
\end{figure}
To understand how TDI works, we have to introduce some notation. Figure \ref{fig1} represents the constellation of LISA schematically. The spacecraft are labelled 1, 2, 3, and they are oriented clockwise. The separating distances are denoted as $\rm L_1, L_2, L_3$, with $\rm L_i$ being the opposite spacecraft, and we will refer with $\rm L'_i$ to the reverse path made by
a photon. primed or unprimed indices refer to 
beams travelling clockwise or counter-clockwise, respectively. The single link response to a GW propagating along $k$ can be written in terms of fractional frequency as:
\begin{equation}
    y_{slr} = \frac{\phi_l(t_s - kR_s(t_s))-\phi_l(t-kR_r(t))}{2(1-kn_l)},
    \label{eq6}
\end{equation}
where index $slr$ stands for ``s''ender, ``l''ink, ``r''eceiver and $\phi_l = n_lh^{ij}n_l$ is the projection of the GW strain $h^{ij}$ on the unit vector of the link associated to the direction $n_l$. The GW strain is referred to the Solar System Barycenter (SSB) frame and can be decomposed as:
\begin{equation}
    h^{\rm SSB}_{ij} = h^{S}_{+}\epsilon^+_{ij} + h^{S}_{\times}\epsilon^\times_{ij},
\end{equation}
where $\epsilon^{+,\times}_{ij}$ are polarization tensors and $h^S_{+,\times}$ are given in the source frame. The vector $R_i$ defines the position of $i$-th spacecraft. The time $t_s$ can be approximated as $t_s = t - |R_r(t)-R_s(t_s)| \sim t - L_l$. With this approximation, we can express Eq.\ref{eq6} as a function of time coordinate $t$:
\begin{equation}
    y_{slr} = \frac{\phi_l(t - kR_s - L_l)-\phi_l(t-kR_r)}{2(1-kn_l)}.
\end{equation}
To study the response of each link (single arm) to the GW signal, it is useful to introduce the delay operator $\mathcal{D}_i$:
\begin{equation}
    \mathcal{D}_iy_{slr}(t) = y_{slr}(t-L_i),
\end{equation}
that we will later express through its shorthand notation $y_{slr,i}$.

The main idea of TDI is to add the delayed signals of LISA together so that the laser frequency noise terms add up to zero. This amounts to seeking data combinations that cancel the laser frequency noise. There exist different versions of TDI, corresponding to different levels of accuracy of the LISA
geometry taken into account. In this work, we adopt the first generation of TDI, that is valid for a static LISA constellation and does not account for rotation and flexing. The first condition emphasizes that all
armlengths are constant in time, while the second condition implies that
the light travel time in one arm is independent of the light propagation
direction, meaning that  $L_i(t) = L_i$ and $L_i = L'_i$. A direct consequence is that the delayed operators commute.

A possible realization of TDI defines three variables $X$, $Y$ and $Z$, corresponding to pairwise  Michelson-like interferometers. As an example, in Eq.\ref{eq10} we report the expression of the $X$ channel in the first generation of TDI, while $Y$ and $Z$ are obtained by cyclic permutation of indices $1 \rightarrow 2 \rightarrow 3 \rightarrow 1$. $X(t)$  can be visualized as the difference of two sums
of measurements, each corresponding to a specific light path from the laser on
board spacecraft 1. 
\begin{equation}
\begin{split}
X(t) =& [(y_{1} + y_{3,2}) + (y_{1,22} + y_{2,322})] \\
&- [(y_1 +y_{2,3}) + (y_{1,33} + y_{3,233})]    
\end{split}
\label{eq10}
\end{equation}
The first square-bracket term in Eq.\ref{eq10} represents a light-beam transmitted from spacecraft 1 and made to bounce over the spacecraft 2 and 3. The second corresponds to another beam also originating from the same laser (equal laser frequency noise), but bouncing off spacecraft 3 first and then spacecraft 2. When they are recombined, they will cancel the laser phase fluctuations exactly, having both experienced the same total delay. Using the new variables, one should extend the definition of PSD in Eq.\ref{eq2} as:
\begin{equation}
    E[\tilde{n}_i(f)\tilde{n}_j^*(f')] = \frac{1}{2}\delta(f-f')S_{n,ij}(f),
\end{equation}
where Latin indices run over the different TDI variables. The instrumental noise that enter the Michelson TDI variables X, Y and Z
are correlated by construction. However, for GW data analysis, it is
more convenient to have data streams with uncorrelated noise. Therefore, we use the optimal TDI variables obtained directly from the Michelson combinations: 
\begin{equation}
\begin{split}
    &A(t) = \frac{Z-X}{\sqrt{2}}\\
    &E(t) = \frac{X-2Y+Z}{\sqrt{6}}\\
    &T(t) = \frac{X+Y+Z}{\sqrt{3}},
\end{split}
\end{equation}
named according to the inventors of TDI, Armstrong (A), Estabrook (E) and
Tinto (T). Since the PSD of the noises are uncorrelated, these variables are also called orthogonal modes. In particular, the T channel has a very low sensitivity to the gravitational wave signal. It provides a ’null’ channel that can be neglected for our analysis. Conversely, the A and
E channels can be seen as two interferometers with a relative orientation of $45^{\circ}$, providing the correspondent measures of the plus and cross polarization of the incident GW.

\section{Review of Kludge waveform families}\label{III}
We turn now to the description of the formalism employed to model the inspiral of EMRIs. Since we are dealing with a binary system of small mass ratio, the gravitational waveform may be obtained accurately using black hole perturbation theory, i.e. by treating the secondary BH as a perturbation to a given black hole background. In particular, one could calculate the gravitational self-force by expanding the background metric in terms of the mass ratio \cite{Sasaki2003}.
The extreme mass ratio also guarantees that the orbital parameters change on a much longer timescale than the orbital period. This implies that the inspiral waveform could be approximated by “snapshot” waveforms, calculated by assuming that the small object is moving along a geodesic. These snapshots are constructed using the Teukolsky equation \cite{Teukolsky1973} which describes the first order change to the curvature tensor of a black hole due to some perturbation. However, Teukolsky-based (TB) waveforms are computationally expensive to generate, as they require the numerical integration of the Teukolsky equation and summation over a large number of modes. These difficulties motivate the construction of approximate families of waveforms that capture the main features of the true signals, at a much lower computational cost. One possible approach is to construct post-Newtonian (PN) {\footnote{PN theory is a systematic approximation method solving the Einstein field equations (and the equations of the motion for a source) in the form of power series of the small parameter $v/c$.}} waveforms, which have the advantage of being analytic and therefore very easy to generate, but cannot fully capture the evolution close to coalescence, where the $v/c$ is ${\cal O}(1)$ invalidating the series expansion approach \cite{Blanchet2014}.
Another approach involves a particular class of approximated waveform, which is known as
kludge, which we will detail in this section. The basic idea behind the kludges is to combine different prescriptions for the orbital evolution and GW emission (not necessarily in a self-consistent way).
Essentially, there exist three kinds of kludges: they are the Analytic Kludge (AK), the numerical Kludge (NK) and the Augmented Analytic Kludge (AAK). 

In general, the kludge class describes the inspiral of a CO, treated as a point mass $\mu$, around a Kerr MBH with mass $M$ and spin $a$. The geometry of the system is depicted in Fig.1 of \cite{Barack2004} and can be described by the following set of parameters.
The direction of the spin for the MBH is represented by the unit vector $\vec S$, or alternatively, the angles $(\theta_K, \phi_K)$. Instead, $\vec L$ represents the direction of the CO's orbital angular momentum, and the angle between $\vec L$ and $\vec S$ is labelled as $\iota$ (inclination). There are two additional angles: the azimuthal angle of $\vec L$ denoted with $\alpha$, and the angle between $\vec L \times \vec S$ and the direction to the pericenter of the CO orbit marked as $\tilde \gamma$.\\
In general, the orbits of COs can be eccentric and non-equatorial. In this contest, $e$ and $p$ represent, respectively, the eccentricity and the semi-latus rectum, from which we obtain the pericenter $r_p = p/(1+e)$ and the apocenter $r_a = p/(1-e)$. For the CO’s motion, $\phi$ is the mean anomaly, and $\nu$ is the correspondent orbital frequency. All these quantities have clear Keplerian meaning for broad orbits; however, we continue using them also in the ultra-relativistic regime close to the separatrix. Further to the parameters describing the orbital motion, there are also the sky location of the source $(\theta_S, \phi_S)$ in the SSB, and the luminosity distance $D_L$, computed directly from the redshift $z$. Thus, to summarise, in the kludges family, 14 parameters are needed to describe the EMRI waveform:
$\{M, \mu, a,p,e,\phi,\alpha,\tilde \gamma, \iota,D_L,\theta _S,\phi _S, \theta _K, \phi _K\}$. Actually, they would be 17, also considering the spin of the secondary object, but its effect on the dynamics and GW emission is small, and it is irrelevant to our analysis.

\subsection{Analytic Kludge}
The AK has been developed by Barack and Cutler \cite{Barack2004}. It approximates the EMRI system as being, at any instant, a Keplerian binary emitting a quadrupolar wave  \cite{Mathews1963}.
 In this waveform, the orbital evolution is given by five first-order ordinary differential equations of $(\phi, \nu, \tilde \gamma, \alpha, e)$. The equations combine several leading order contributions from PN (Post-Newtonian) dynamics.
 For $\nu$ and $e$, the expressions are accurately through 3.5PN $\tilde \gamma$ and $\alpha$ equations are accurate through 2PN order. The PN expressions are expressed by equations (28-31) in \cite{Barack2004}.
The EMRI orbital elements are  integrated between $t=0$ and the observed time $T_{obs}$. If the EMRI plunges during the observation, the inspiral evolution has to be truncated and this can be done in two different ways. The first consists in the "Schwarzschild" analytic kludge which uses the last stable orbit for non-spinning MBH as a proxy for the plunge.  Alternatively, the "Kerr" Analytic Kludge takes into account the MBH spin information, identifying the last stable orbit  defined by a Kerr separatrix. The choice of different plunging conditions defines how close to MBH horizon we can approach and, therefore, the  bandwidth of the signal and its SNR.  The Schwarzschild version of AK underestimates signal's strength, while the Kerr version strongly overestimates it as PN equations approximate poorly dynamics near  the horizon.

The AK waveform is very efficient from a computational point of view, and for that reason, it is widely used to study EMRIs. However, its poor accuracy does not allow us to represent the expected signal.

\subsection{Numerical Kludge}
The NK waveform has been developed by \cite{Babak2008}, and the idea behind it is to combine the accurate particle trajectory to an approximate expression for the GW emission.
The first step in constructing a NK waveform is to compute the trajectory that the inspiralling body follows in the Boyer-Lindquist coordinates of the Kerr spacetime of the central black hole. Bound black hole orbits (in the absence of GW radiation) admit four constants of the motion, allowing us to rewrite the geodesic equations as a system of first order differential equations.  These four constants are: the rest mass, the energy $E$, the axial angular momentum $L_z$ and the Carter constant $Q$ \cite{Carter:1968}.  The first order equations can be written in the following form:
\begin{equation}
    \begin{split}
        &\frac{dr}{d\tau} = \pm \sqrt{V_r},\\
        &\frac{d\theta}{d\tau} = \pm \sqrt{V_\theta},\\
        &\frac{d\phi}{d\tau} = V_\phi,\\
        &\frac{dt}{d\tau} = V_t.
    \end{split}
    \label{eq15}
\end{equation}
Solutions of the geodesic equations (Eq.\ref{eq15}) are uniquely
determined if we specify $E$, $L_z$ and $Q$ and initial orbital position. Often these quantities are expressed in terms of $(r_a, r_p, \theta_{min})$: from the roots of $V_r$ we can determine the periastron $r_p$ and the apoastron $r_a$ and, following their definition, we arrive to compute the semi-latus rectum and the eccentricity. The angle $\theta_{min}$ is instead determined by the root  with the smallest value of $V_\theta$ (turning point).
Exploiting $Q$ and $L_z$, we can obtain the inclination angle $\iota$, as $\cos \iota = L_z / \sqrt{Q + L_z^2}$. The geodesic motion has to be augmented with a dissipation due to GWs, which (in adiabatic approximation) is described by orbit average changes on the orbital constants. Then, (Eq.\ref{eq15}) are integrated to obtain the trajectory of the CO. Finally, the waveform can be obtained from the inspiral trajectory using quadrupole (or quadrupole-octupole) approximation as it were in the flat space-time.\\
The NK waveform provides a good agreement with the numerical Teukolsky-based waveform, thus superseding the AK. The model, however, is computationally more expensive since it requires the integration of the trajectory both in phase and coordinate space.

\subsection{Augmented Analytic Kludge}
The AK model can be up to 15 times faster \cite{Chua2017} than the NK model at generating waveforms. This efficiency could increase for longer integrations, but not at higher eccentricity because more harmonics are needed in the AK approximation. However, AK waveforms suffer, already at the early-inspiral stage, from dephasing with respect to NK waveforms  due to the mismatched frequencies in the two models. The AAK model was first introduced in \cite{Chua2015} to cure this issue. The AAK waveform uses information from the NK model to improve the faithfulness of AK waveforms without significantly increasing their computational cost. The main idea is
to extend parameters of AK model beyond their physical meaning to match the frequencies of NK waveforms. AAK generates a small section of trajectory with NK, and then maps the AK trajectory to the NK result and finds out the best-fit parameters. Let us give a more detailed description of this procedure.

A bound geodesic orbit in Kerr spacetime is characterized by three fundamental frequencies $\Omega_{r,\theta,\phi}$ for the radial, polar and azimuthal components of the motion. These frequencies take a simple form, choosing a timelike parameter $\lambda = \int d\tau/\Sigma$ (the so-called \textit{Mino time}). Indeed, the frequencies are given by:
\begin{equation}
    \begin{split}
        &\Omega_r = \frac{2\pi}{\Lambda_r \Gamma},\\
        &\Omega_\theta = \frac{2\pi}{\Lambda_\theta \Gamma},\\
        &\Omega_\phi = \lim _{N \rightarrow \infty} \frac{1}{N^2\Lambda_r \Lambda_\theta \Gamma}\int ^{N\Lambda_r}_0 d\lambda_r \int ^{N\Lambda_\theta}_{0} V_\phi\, d\lambda_\theta ,
    \end{split}
\end{equation}
where $\Lambda_{r,\theta}$ represent the radial and polar period, and $\Gamma = <dt/d\lambda>$ the average rate (in analogy with the Lorentz factor):
\begin{equation}
    \begin{split}
        &\Lambda_r = 2\int^{r_a}_{r_p} \frac{dr}{\sqrt{V_r}},\\
        &\Lambda_\theta = 4\int^{\pi/2}_{\theta_{min}}\frac{d\theta}{V_\theta},\\
        &\Gamma = \lim _{N \rightarrow \infty} \frac{1}{N^2\Lambda_r \Lambda_\theta}\int ^{N\Lambda_r}_0 d\lambda_r \int ^{N\Lambda_\theta}_{0} V_t\, d\lambda_\theta ,
    \end{split}
\end{equation}
In terms of the fundamental frequencies, the periapsis and Lense-Thirring precession rates are given by $\Omega _{pre} = \Omega_\phi -\Omega_r$
and $\Omega_{LT} = \Omega_\phi - \Omega_\theta$, respectively. We can introduce for a BH with mass $M$ the dimensional frequencies $\omega _{r,\theta, \phi} = \frac{\Omega _{r,\theta, \phi}}{2\pi M}$, which in the Newtonian limit tend to the orbital frequency, i.e. $f_{orb} = \omega_r = \omega_\theta = \omega_\phi$. In this limit, the periapsis and Lense-Thirring precession are zero, but in AK they are introduced  manually through the PN equation. In particular, the precession frequency would be $f_{pre} = \dot \gamma + \dot \alpha$ and Lense-Thirring frequency $f_{LT} = \dot \alpha$. Thus, we can match the three frequencies in the AK and in the Kerr formalism, producing an endomorphism over the AK parameter space. Typically, the frequencies are expressed in terms of $(M, a, p)$, but also the set $(e, \iota, p)$ can be used. For example, given $(M, a, p)$ of a BH, the match between frequencies is done by introducing the 
non-physical values $(\tilde M, \tilde a, \tilde p)$ as:
\begin{equation}
    \begin{split}
        f_{orb} =& \dot \phi(\tilde M, \tilde a, \tilde p) = \omega_r(M, a, p),\\
        f_{LT} =& \dot \alpha (\tilde M, \tilde a, \tilde p) = \omega_\theta (M, a, p) - \omega_\phi (M, a, p),\\
        f_{pre} =& \dot \gamma(\tilde M, \tilde a, \tilde p) + \dot \alpha(\tilde M, \tilde a, \tilde p) \\
        =& \omega_\phi(M, a, p) - \omega_r (M, a, p).
    \end{split}
    \label{eq18}
\end{equation}
Substituting the parameters $\tilde M, \tilde a, \tilde p$ with $(M, a, p)$  in the AK model provides a correction of its frequencies along the entire inspiral trajectory. Next, the waveform can be generated using the AK framework with  the improved orbital motion. 

An improved AAK model is implemented by \cite{Chua:2020} as a part of the package \textit{FastEmriWaveform} (FEW). The "classic" AAK described above, builds a trajectory by using the frequency evolution from the Numerical Kludge and mapping it onto the frequency basis. The new improved AAK, instead, computes directly the fundamental frequencies exploiting a 5PN trajectory developed in \cite{Fujita2020}.
Finally, a fast method of plunge handling has been added to the AAK implementation. In general, the CO plunges when its orbit along the phase-space trajectory becomes unstable (separatrix):
\begin{equation}
\begin{split}
\frac{\partial^2 V_r(r,a,E,L_z,Q)}{\partial^2 r} \leq& \frac{\partial V_r(r,a,E,L_z,Q)}{\partial r} \\
=& V_r(r,a,E,L_z,Q) = 0.
\end{split}
\end{equation}
The separatrix, in general, depends upon three parameters ($e$, $p$, $\iota$). In literature, orbits along the separatrix with $e = 0$ are referred to as the innermost stable spherical orbit (ISSO). If $\iota =  0$ this orbit is usually called the innermost stable circular orbit (ISCO) instead. In the FEW framework, the trajectory has been computed until it reaches 0.1 of the separatrix, computed numerically as shown in \cite{Sten2020}.
Noteworthy, another EMRI waveform has been embedded in FEW; it is a fully relativistic waveform based on the interpolated self-force calculations. However, this model covers only non-spinning BHs.

\section{From EMRI Population to GWB }\label{IV}
In order to provide astrophysical motivated estimates of the GWB generated by a cosmic population of EMRIs, we consider several models presented in \cite{Babak2017}, reported in Table \ref{tab:tab1} for completeness.
%%%%%%%%%%%%%%%%%%%%%
\begin{table*}
\centering
\begin{tabular}{ccccccc|ccc}
\hline
 & Mass &  MBH & Cusp & $M$--$\sigma$ & & CO & & EMRI rate [$\mathrm{yr}^{-1}$] & \vspace{-0.04in}\\
Model &  function &  spin &  erosion &  relation & $N_\mathrm{p}$ &  mass [$\rm M_odot$] & Total  & Detected (AKK)  & Detected (AKS) \\
% &   &   &   &   &  &   &   &  (AKK) &  (AKS)\\
\hline
M1 & Barausse12 & a98   & yes & Gultekin09    & 10  & 10 & 1600 &  294& 189\\ %def
M2 & Barausse12 & a98   & yes & KormendyHo13  & 10  & 10 & 1400 &  220& 146\\ %Msigmapess
M3 & Barausse12 & a98   & yes & GrahamScott13 & 10  & 10 & 2770 &  809& 440\\ %Msigmaopt
M4 & Barausse12 & a98   & yes & Gultekin09    & 10  & 30 &  520 & 260 &221\\ %mCO30
M5 & Gair10     & a98   & no  & Gultekin09    & 10  & 10 &  140 &  47& 15\\ %MFpess
M6 & Barausse12 & a98   & no  & Gultekin09    & 10  & 10 & 2080 &  479& 261\\ %NOerosion
M7 & Barausse12 & a98   & yes & Gultekin09    & 0   & 10 & 15800 &  2712& 1765\\ %NOplunges
M8 & Barausse12 & a98   & yes & Gultekin09    & 100 & 10 &  180 &  35& 24\\ %100plunges
M9 & Barausse12 & aflat & yes & Gultekin09    & 10  & 10 & 1530 &  217& 177\\ %aflat
M10 & Barausse12 & a0    & yes & Gultekin09    & 10  & 10 & 1520 &  188& 188\\ %a0
M11 & Gair10     & a0    & no  & Gultekin09    & 100 & 10 &   13 &  1& 1\\ %pess 
M12 & Barausse12 & a98   & no  & Gultekin09    & 0   & 10 & 20000 &  4219& 2279\\ %opt
\hline
\end{tabular}
\caption{List of EMRI models taken from \cite{Babak2017} and considered in this work. Column 1 defines the label of each model. For each model the following quantities are specified: the MBH mass function (column 2), the MBH spin model (column 3), whether or not the effect of cusp erosion is included (column 4), the $M-\sigma$ relation (column 5), the ratio of plunges to EMRIs (column 6), the mass of the COs (column 7), the total EMRI merger rate ($\rm yr^{-1}$) up to $z = 4.5$ (column8). In column 9 and 10 the detected EMRI rate per year is reported for two different variants of  AK waveforms (AKS and AKK truncate the waveform respectively at Schwarzschild and Kerr ISCO). The full definition and implementation of each ingredient entering the models can be found in \cite{Babak2017}.}
\label{tab:tab1}
\end{table*}
%%%%%%%%%%%%%%%%%%%%%

The main idea behind the 12 different models was to bracket the expected range of EMRI rates spanning the wide range of uncertainties affecting EMRI formation, including: the cosmic evolution of the MBH mass function, the relation between MBH mass and density of the surrounding stellar environment, the impact of core scouring following galaxy mergers and so on. For each model, the distribution of cosmic EMRIs was built, and mock catalogues were generated  by Monte Carlo sampling from the distribution. In particular, 10 Monte-Carlo realizations of the expected population of EMRIs plunging within 1 year were generated. Putting together, they consist in a library which includes all EMRI events occurring in the Universe during 10 years for each of the 12 models. Every catalogue provides the primary mass $M$, the redshift of the event $z$, and MBH spin $a$ of each event. For the EMRIs it has been considered a secondary mass of $10 M_{\odot}$ or $30 M_{\odot}$ as shown in the table. The expected event rate (number of sources per year) varies across models 
 from tens to tens of thousands. This large range reflects the astrophysical uncertainties on the model parameters, e.g.: the choice of the number of direct plunges or the MBH mass function. Full details about the models can be found in \cite{Babak2017}.
The GW signal modelling requires 14 parameters listed in section \ref{III}. For each EMRI inside the catalogs, we associate the cosine of inclination angle $\iota$ by drawing it randomly from a uniform distribution between $[-1,1]$ since we consider that EMRIs form in a spherical bulge and not in a disk-like structure. Note that $\iota$ spans in the interval $[0,\pi]$ and that prograde orbit are associated  for $0 \leq \iota\leq \pi/2$, while retrograde orbit for $\pi/2 \leq \iota\leq \pi$. 
To determine the initial eccentricity and the semi-latus, we exploit \cite{Babak2017}, which has shown that, evolving large samples of COs, the eccentricity distribution at plunge is nearly flat in the range $0<e_{\rm pl}<0.2$. Then, we use the same strategy adopted in \cite{Bonetti2020}:
\begin{itemize}
    \item for each event in the catalogs we draw  the eccentricity at the last stable orbit $e_p$ in the range $[0;0,2]$. Since we already know $M$ and the spin $a$, we can also estimate the radius at the last stable orbit.
    \item we integrate the orbital elements of the event backwards in time for $T_{\rm back}$ years (i.e. using the orbit-averaged equations in \cite{Peters1964}).
    \item we draw  $N_{\rm back} = \textrm{int}(T_{\rm back}/10)$ points randomly in the range $[0;T_{\rm back}]$ in order to select different evolutionary points of each EMRI. Here, the division by 10 is made because, as we mentioned, we collect 10 catalogs of EMRIs for every model, each representing one year of observation. $T_{\rm back}$ is computed as $T_{\rm back} = 20(\frac{M}{10^4 \rm M_{\odot}})\rm yr$, because the time taken to cover the same amount of gravitational radii scales linearly with MBH mass. Thus, EMRIs with low mass primarily emit in the LISA band only over the last years of their inspiraling, while EMRIs with larger MBH masses emit in the LISA band for much longer prior to the plunge;
    \item for each $N_{\rm back}$ we collect the semi-major axis $a_{\rm sm,0}$  and the eccentricity $e_0$ at the corresponding backward time. Following the definition of semi-latus, we compute it as $p_0 = a_{\rm sm,0}(1-e_0^2)$.
\end{itemize}

With this procedure we are creating $N_{\rm back}$ copies of each EMRI from the catalogs of \cite{Babak2017} with same redshift, masses, spin and inclination angles. Nevertheless, the procedure is not expected to introduce any bias in the computation of the background because the catalogs provide a smooth coverage of the relevant MBH mass redshift and spin range. 

Concerning the sky position $(\theta_S,\phi_S)$ and spin orientation $(\theta_K,\phi_K)$, we assume that the corresponding unit vectors are isotropically distributed on a sphere. Finally, the three initial phases corresponding to the orbital phase, the precession phase of the periapsis and the precession phase  of the orbital plane (Lense-Thirring) are uniformly distributed between 0 and $2\pi$. To compute the waveform with AAK we convert them in terms of radial, polar and azimuthal phases $(\Phi_{r,0}, \Phi_{\theta,0}, \Phi_{\phi,0})$ using the system analogues to Eq.\ref{eq18}.

\subsection{Definition of the catalogues} \label{sec4a}
After completing the list of parameters, we compute the signal of each EMRI. Since we want to evaluate the power spectrum of the unresolved GW background in the LISA detector, we have to work in frequency domain using the optimal combination of TDI described in Sec. \ref{sec2.b}. However, the AAK code returns the signal in time domain. To inject the signal in the detector, we therefore perform the following steps:
\begin{enumerate}
    \item Computation of $h_+(t), h_\times(t)$ with AAK + 5PN trajectory, considering an observation time of 4 years and a measuring cadence of $15s$.
    \item Computation of the optimal TDI combination $A(t), E(t)$.
    \item Computation of their Fourier transforms $\tilde{A}(f), \tilde{E}(f)$.
\end{enumerate}
This procedure has a variable computational cost depending on the initial parameters of the source. For example, a source with high eccentricity $e_0$ takes more time because the number of harmonics needed in the computation is higher.

\begin{table}
\begin{tabular}{ccccccc}
\hline\hline
Model& $N_{\rm tot}$& $N_{f}$& SNR$_{\rm tot}$ & SNR$_{\rho >1}$ & SNR$_{f}$ &Detections\\
\hline
M1& 1225158  & 31764& 571& 534& 460& 366\\
M2& 580149& 17030& 434& 418& 352& 386\\
M3& 2030059& 68100& 1174& 1121& 839& 1123\\
M4& 43607& 11166& 960& 926& 750& 1713\\
M5& 1772409& 44011& 727& 695& 535& 622\\
M6& 396800& 2486& 48& 44& 27& 35\\
M7& 8218425& 303348& 4940& 4748& 3960& 4619\\
M8& 89010& 3620& 63& 61& 51& 53\\
M9& 872231& 31356& 501& 480& 419& 426\\
M10& 823589& 30597& 478& 459& 408& 406\\
M11& 34724& 287& 4.53& 4.11& 2.32& 0\\
M12& 16547658& 394583& 6475& 6200& 4553& 5580\\
\hline\hline
\end{tabular}
\caption{The effect of source selection. Columns 2 and 3 are the initial and final number in the EMRIs catalogues before and after the source selection process. Column 4 is the original SNR of GWB calculated using the simplified version of the AK waveform assuming 4 years of observation time. Column 5 is the SNR after removing sources with SNR$<1$. Column 6 is the final SNR computed after removing also sources with $e_0 > 0.9$ or $p_0 < 10$. The last column corresponds to the number of detected EMRIs, defined as those with SNR$>20$.}
\label{tab1}
\end{table}

\begin{figure}
    \centering
    \includegraphics[scale=0.4]{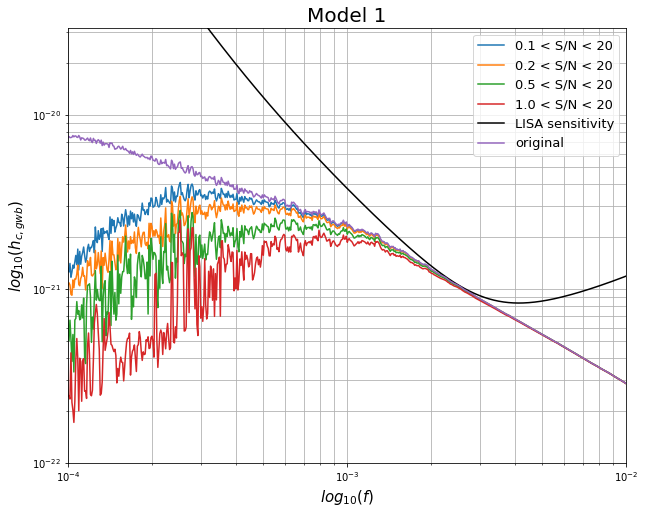}
    \caption{Characteristic strain of the GWB generated for model M1 listed in table  \ref{tab:tab1}. The  colors refer to different SNR thresholds for the EMRI included in the computation, as labelled in figure. The gravitational signal has been computed using a simplified, inclination-polarization averaged version of the AK waveform.}
    \label{fig3}
\end{figure}

The whole procedure takes on average $\sim 120$ seconds per source, which is a non-negligible computational time, given that the populations we built are composed of a number of sources which spans from tens of thousands up to several millions.
For instance, model M12 counts almost 16 million EMRIs, meaning the background evaluation would take about 60 years on a single CPU. We, therefore, need to find a strategy to speed up the process. A possibility is to select only the sources that produce a significant contribution to the GWB. For this purpose, we evaluate a preliminary SNR for each EMRI exploiting
the inclination-polarization averaged version of the AK waveform, following the same approach as in \cite{Bonetti2020} and truncating the EMRI evolution at the Schwarzschild last stable orbit. Then, we introduce a cut in the population, removing all the sources below a chosen SNR threshold. Finally, we compute what fraction of the total SNR is lost due to the cut. We tried different cuts, and we decided to fix the threshold to $\rho = 1$, which allows removing the greatest number of sources without losing too much total SNR. After removing the weak sources, we observe an SNR loss of $\sim 5\%$ only (cf. columns 4 and 5 of table \ref{tab1}). Fig.\ref{fig3} shows an example of this simplified GWB computation for model M1 for different SNR cuts. The selection affects the GWB mostly at $f< 1\rm mHz$, where LISA starts to lose sensitivity. Conversely, where the instrumental curve is closer to the EMRI GWB level, the backgrounds are similar to each other. 

\begin{figure}
    \centering
    \includegraphics[scale=0.45]{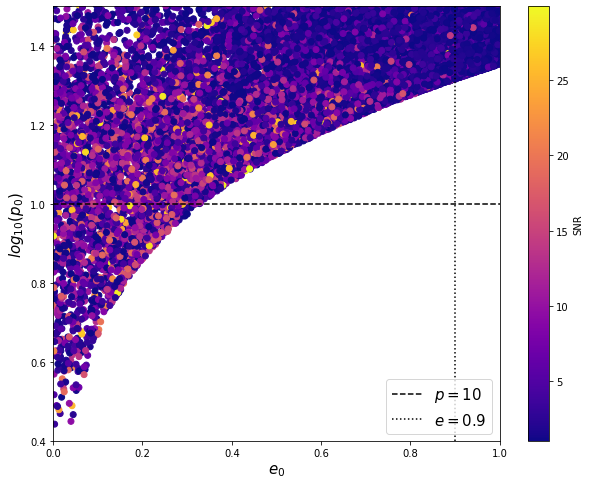}
    \caption{Distribution of EMRIs in the plane $e$-$\log_{10}(p)$ for model M1. The color scale refers to the SNR of EMRIs computed through the simplified AK waveform. Note that louder EMRIs are generally characterized by smaller $p_0$ rather being very eccentric.}
    \label{fig4}
\end{figure}

The choice of the AAK waveform imposes two further selection criteria: (i) a cut in eccentricity and (ii) a cut in semi-latus rectum. 
The number of harmonics needed to construct the waveform, and consequently the computational cost, is a steep function of eccentricity.
We verified that for sources with eccentricity bigger than 0.9, the waveform code does not terminate the computation, and therefore we cut from the catalog all sources with $e_0>0.9$.
Moreover, we also remove the sources with initial semi-latus smaller than 10 since the global fit for the inspiral turns out to be difficult for such tight EMRIs. The cut in semi-latus removes only $\sim 1\%$ of the sources from the original catalogs, as reported in the third column of table \ref{tab2}.
To summarize, the final EMRIs catalogs  are composed by sources with $e_0 < 0.9$, $p_0 > 10$ and individual SNR greater than 1, calculated with a simplified AK waveform.
\begin{table}
    \centering
    \begin{tabular}{ccc}
    \hline\hline
         Model& $\%_{e>0.9}$ & $\%_{p<10}$ \\
         \hline
         M1 & 26.5 & 1.11 \\
         M2 & 30.3 & 0.92 \\
         M3 & 28.0 & 0.84 \\
         M4 & 26.8 & 1.08 \\
         M5 & 22.2 & 1.17 \\
         M6 & 10.3 & 1.56\\
         M7 & 26.5 & 1.11 \\
         M8 & 28.1 & 1.12 \\
         M9 & 26.3 & 1.14 \\
         M10 & 26.8 & 1.07 \\
         M11 & 11.1& 1.35 \\
         M12 & 21.6 & 1.15\\
         \hline\hline
    \end{tabular}
    \caption{Percentage of EMRIs removed due to the selection procedure aiming at speeding-up the GWB evaluation. Columns 2 refers to the sources subtracted with $e_0>0.9$, while Column 3 are the fraction of sources with $p_0<10$.}
    \label{tab2}
\end{table}
It is important to notice that cutting in eccentricity and the semi-latus rectum induced the loss of a significant part of the original background information. As reported in column 6 of table \ref{tab1}, the final SNR, after taking into account this further selection, is about $20-30\%$ lower compared to the full population. Despite the number of removed EMRIs at high eccentricity is larger, most of the signal loss is due to the selection in the semi-latus rectum. Indeed, we found that the percentages of SNR lost due to semi-latus selections are greater than the other two, as seen in Tab.\ref{tab5}. Fig.~\ref{fig4} represents the distribution of the sources in the plane eccentricity/semi-latus for model M1. First of all, we observe that $e>0.9\,\cap\,p<10 = \emptyset$, and the two cuts are independent. The gradient of colour shows that the $p<10$ region host  multiple sources with single SNR between 10 and 20, which in principle could contribute to increasing the final background, whereas only low SNR sources are found at $e>0.9$. 

Despite the above-mentioned caveats, we stress once again the efficiency of this selection: we are able to maintain almost $80\%$ of the original signal just considering on average $3\%$ of the sources in the catalogs, making the computation cost manageable \footnote{Notice that the computational cost is still non-negligible. Taking, for instance, M12, the computation of all its sources after the selections requires almost one week exploiting 100CPU}. 

\begin{table}
    \centering
    \begin{tabular}{ccccc}
    \hline\hline
         Model& $\% \rm SNR_{tot}$ & $\% \rm SNR_{\rho}$ & $\% \rm SNR_{p}$ & $\% \rm SNR_{e}$\\
         \hline
         M1 & 20 & 7 & 11 & 2 \\
         M2 & 18 & 4 & 11 & 3 \\
         M3 & 28 & 5 & 19 & 4 \\
         M4 & 22 & 4 & 13 & 6 \\
         M5 & 26 & 4 & 18 & 4 \\
         M6 & 43 & 8 & 30 & 5\\
         M7 & 20 & 4 & 11 & 5 \\
         M8 & 16 & 3 & 12 & 1 \\
         M9 & 16 & 4 & 9 & 3 \\
         M10 & 14 & 4 & 8 & 2 \\
         M11 & 49 & 9 & 39 & 1\\
         M12 & 29 & 4 & 18 & 7\\
         \hline\hline
    \end{tabular}
    \caption{Percentage of SNR after different sources selections. Column 2 is the total percentage lost, column 3 are referred after removing the dimmest source with $\rm SNR < 1 $, while column 4 and 5 are respectively the SNR after deleting sources with $p<10$ and $e>0.9$. }
    \label{tab5}
\end{table}

\subsection{Background Computation}
\begin{figure}
    \centering
    \includegraphics[scale=0.6]{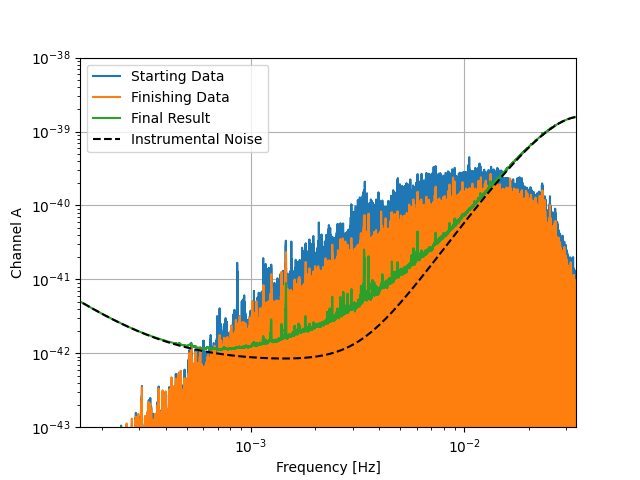}
    \caption{Illustration of the procedure for estimating the unresolved GWB from EMRIs. This plot refers to M12, and it takes 4 iterations for the algorithm to converge. The starting data are represented in blue, while the orange shows the result at the end of the procedure. The final combined instrument plus median confusion noise is represented in green.}
    \label{ife}
\end{figure}

Using the population described in the previous section, we generate the AAK waveform for each EMRI, inject them in the optimal LISA TDI variables, and we analyze the data directly on those TDI data streams. For each model, we compute the initial background as the sum of all the signals for both channels A and E (in the frequency domain). We should remove the sources whose SNR exceeds a set threshold to obtain the final confusion noise.  The detection threshold is adjusted for the different source types, taking into account search and parameter estimation studies. The Mock LISA Data Challenge results suggest that EMRIs with SNR as low as $\rho_0 =20$ could be identified \cite{Babak2010}. The SNR for a given source is defined as:
\begin{equation}
    \rho = \sqrt{\sum _k (h_K|h_K)},
\end{equation}
where $K$ refers to the optimal TDI variables $(A, E)$ and the inner product between $a$ and $b$ is given by
\begin{equation}
    (a|b) = 2 \int ^\infty _0 [a^*(f)b(f) + a(f)b^*(f)]/S_n(f),
\end{equation}
where $S_n(f)$ is the one-sided noise PSD, which is $S_K(f)$ for the different TDI variables. However, in presence of a GWB the PSD becomes $S_n(f) = S_{\rm instr}(f) + S_{\rm gwb}(f)$, taking into account both the instrumental noise, $S_{\rm instr}(f)$, and the confusion noise, $S_{\rm gwb}(f)$. To solve this problem, we use the Iterative Foreground Estimation (IFE) algorithm, developed by \cite{Karnesis:2020}, which is based on an iterative process to evaluate $S_n(f)$, whose main steps are the following:
\begin{enumerate}
    \item After generating the total signal, $S_{\rm gwb}(f)$, from the full sample of sources, the code computes the SNR of each source considering only the LISA instrumental noise. We call it SNR in isolation $\rho _i ^{\rm iso}$.
    \item The code evaluate the new PSD, $S_{n,k}(f)$, as the sum of $S_{\rm instr}(f)$ and the first estimation of $S_{\rm gwb}(f)$. Then, the algorithm computes the new SNR $\rho _i$, using $S_{n,k}(f)$ ($k$ refers to the iteration number). 
    \item If $\rho _{i} > \rho _0$, the code subtracts the source from the confusion noise $S_{\rm gwb}(f)$. Instead, if $\rho _i ^{iso} < \chi \rho_0$, the program skips the computation of the SNR for that source and adds it automatically to the final foreground. Here, $\chi$ is a prefixed factor, typically smaller than one. 
    \item After subtracting the bright sources in the previous step, the algorithm returns to the step 2., estimating $S_{n,k+1}(f)$ and iteratively performing the same procedure.
\end{enumerate}
The iteration proceeds until there are no more sources with $\rho _{i} > \rho _0$ or alternatively when $S_{n,k+1}(f) \sim S_{n,k}(f)$, within a fractional tolerance limit. Typically, the algorithm requires 1 to 5 iterations to converge. An illustration of the procedure is represented in Fig.\ref{ife}. 

\section{Results}\label{V}
\begin{figure}
    \centering
    \includegraphics[scale=0.6]{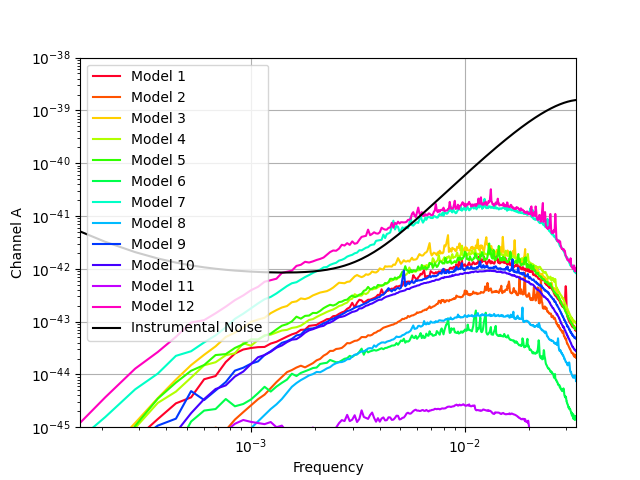}
    \centering
    \includegraphics[scale=0.6]{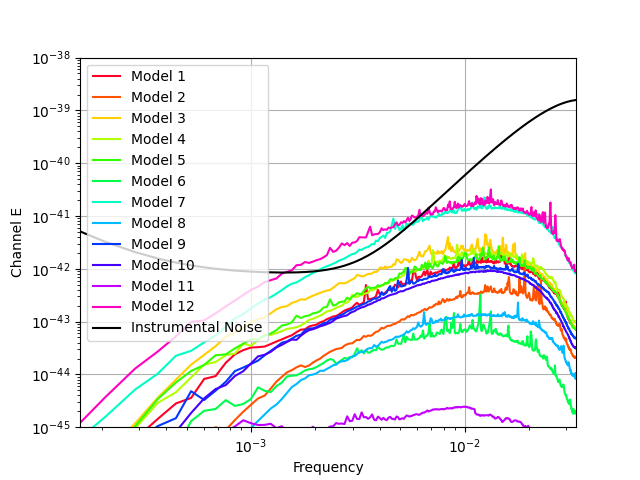}
    \caption{The GWB generated in the 12 different EMRI formation scenarios represented in the TDI variables A (top panel) and E (bottom panel). The black line is the LISA noise in the correspondent channel.}
    \label{fig5}
\end{figure}

Combining the EMRI waveform modelisation, the LISA TDI and the GWB computation, we evaluated the EMRI background for the 12 catalogs  as shown in Fig.\ref{fig5}. The GWB curves for all models lie between the optimistic and the pessimistic scenarios, which are models M12 and M11, respectively.
Uncertainties in the EMRI GWB estimate span about 3 orders of magnitude, which is consistent with the uncertainty in the EMRI rates reported by \cite{Babak2017}. One thing is noteworthy: part of the investigated models predicts a GWB comparable to or higher than the LISA instrumental noise, which therefore cannot be neglected when considering the detectability of other sources \cite{Bonetti2020}. M12 and M7 overcome the LISA level, while M1, M3, M4, M5, M9, M10 are less than an order of magnitude apart from the black line in the bucket, thus making a non-negligible contribution to the total PSD. 

To characterize each single GWB, we compute its SNR. Since we are neglecting the T channel, the overall  SNR  is just the square root of the sum in quadrature of the A and E channels. The SNR of GWB in A and E TDI LISA channel has been computed through \cite{Romano2017,Caprini2019}:
\begin{equation}
    SNR_{i} = \sqrt{T_{\rm obs}\int^\infty_0 df \left( \frac{S_{\rm gwb,i}(f)}{S_{n,i}(f)}\right)^2}
\end{equation}
where $i$ can be A or E and $S_{\rm {GWB},i}$ is the PSD due to the GWB.
Numbers are reported in Table~\ref{tab3} assuming 4 years of data collection. For most models, the SNR of the unresolved GWB is larger than 100, making the signal easily detectable. The GWB produced by models M2, M5 and M8 has a moderate SNR and might be hard to detect, whereas model M11 does not produce an appreciable signal. 
\begin{table}
    \centering
    \begin{tabular}{ccc}
    \hline\hline
    Model& Detections & SNR$_{\rm GWB}$\\
    \hline
    M1& 139& 180\\
    M2& 42& 40\\
    M3& 346& 441\\
    M4& 516& 235\\
    M5& 188& 252\\
    M6& 13& 21\\
    M7& 724& 980\\
    M8& 19& 22\\
    M9& 108& 160\\
    M10& 97& 136\\
    M11& 0& 1.44\\
    M12& 891& 1146\\
    \hline\hline
    \end{tabular}
    \caption{Number of resolvable sources and residual GWB SNR for the investigated astrophysical models. The results have been obtained considering 4 years of observation and using the AAK waveform \cite{Chua:2020}} 
    \label{tab3}
\end{table}

\subsection{Comparison with the literature}
We can perform a comparison between our results and the findings of \cite{Bonetti2020} by matching the SNRs shown in table \ref{tab3} with those in the second last column of table \ref{tab1}. The latter are computed following the same approach as \cite{Bonetti2020}, but applying the same sources selection described in Sec.~\ref{sec4a}. We observe that our SNRs  are lower with respect to those of \cite{Bonetti2020} by a factor between two and four, depending on the model. This difference may be due to some discrepancies between the two approaches used to estimate the EMRI GWB.

The first obvious reason can be ascribed to the choice of the waveform. Indeed, we perform our computation using the innovative AAK, which is more accurate (at the level of the numerical kludge) with respect to the AK waveform, used in \cite{Bonetti2020}. As shown in figure 12 of \cite{Bonetti2020}, the AK model captures the salient features of the waveform, but is less precise in modelling the spectrum close to the final plunge, because it overestimates the frequency of the last stable orbit. This effect is enhanced when truncating the inspiraling at Kerr separatrix (green line), but it is also present in Schwarzschild case. Thus, the excess power in the AK signal at higher frequencies can possibly induce a boost in the SNR. 

A second cause contributing to the discrepancy can  be related to the computation of the SNR itself. Indeed, in our study, the final SNRs are computed including in the noise spectral density both the instrumental and the EMRIs background component, unlike table\ref{tab1} (and \cite{Bonetti2020}) where the astrophysical noise has been neglected. Consequently, the SNR can be lower since the noise level is higher. Moreover, neglecting the EMRI GWB in the background computation can lead to an incorrect subtraction of the resolvable sources.

Finally, the GWB computation in \cite{Bonetti2020} is based on the use of inclination-polarization averaged fluxes, while we automatically take into account the inclination of each individual system with respect to the observation frame injecting the signal in the TDI variables, which ultimately makes our estimation more realistic.

In Table~\ref{tab3}, we also report the number of resolvable sources which have been subtracted from the background using the iterative algorithm. We compare our results 
to the only other two estimates found in the literature, in 
\cite{Bonetti2020} and \cite{Babak2017}. 
In the last column of Table~\ref{tab1}, we report the EMRI detected in 4 years obtained with the same procedure as in \cite{Bonetti2020}, but adjusted with respect to our selection. Our detection rates are more than a factor of three smaller, except for M11, for which we confirm that no EMRI can be resolved. As mentioned, this discrepancy is twofold: (i) due to the AK waveform employed and on whether it is truncated at the Kerr or Schwartzschild last stable orbit,
the truncation at the Kerr separatrix causes an excess power at high frequencies compared to the AAK waveform, inducing a boost in the SNR;
(ii) the instrumental noise level assumed in \cite{Babak2017} is about 1.5 lower (above 2-3 mHz) compared to the currently adopted.
In addition, some of the removed EMRIs, especially those with $p_0<10$, might be sufficiently loud to be resolved (without affecting the GWB). Still, when \cite{Bonetti2020} add (in an approximate fashion) the corresponding EMRI GWB to the LISA sensitivity curve, we find rates of the same order (i.e. column labeled 'AKSb' in table II of \cite{Bonetti2020}).

Compared to numbers obtained by \cite{Babak2017} and reported in table \ref{tab:tab1}, our detection rates are even lower. For example, considering the optimistic case M12, the EMRI detection rate reported in table \ref{tab:tab1} is 4219 per year in the Kerr case and 2279 per year in the Schwartzschild case, while we obtain only 891 detections in 4 years, i.e. only $\approx 200$ events per year. The difference can be ascribed to a number of causes: the selection of the sources in our catalogs, the choice of the waveform, the LISA curve employed at that time and how the resolvable sources are subtracted. Clearly, part of the difference is partially due to the selection of the sources in our catalogs,  the choice of the waveform, and  how the resolvable sources are subtracted, as explained in the previous comparison with \cite{Bonetti2020}. A direct comparison with \cite{Babak2017} is unfortunately not straightforward: \cite{Babak2017} considers catalogs of EMRIs plunging over 10 years and  compute SNR by considering only the last two years before the plunge (integrating backwards in time). They then divide the obtained numbers by 10 to get the detection rates per year. This procedure implies that the EMRI detection rate can somehow be independent of the mission duration, however, as shown in \cite{Bonetti2020}, this does not seem to be the case. The EMRI detection rate indeed depends non-linearly on the observation time, making any direct comparison flawed.

\section{Conclusions and Outlook}\label{VI}
In this work, we produced the first realistic calculation of the signal imprinted in the LISA detector by a stochastic GWB produced by a cosmological population of EMRIs. Three main ingredients have been combined to achieve the goal: the generation of a realistic population of EMRIs, the simulation of their GW signals and their injection in LISA data stream. 
%The procedure followed implies that our estimate of the EMRI GWB is likely the most sophisticated and accurate to date.

We first built the EMRI population from the astrophysical models developed in \cite{Babak2017}, which include the most relevant ingredients driving EMRI formation in the standard relaxation-driven channel. Following the procedure devised in \cite{Bonetti2020}, we constructed from these models the corresponding EMRI catalogs.
To speed up the foreground computation, we removed the sources contributing less to the final GWB by putting a minimum threshold in the single SNR pre-computed with a simplified, inclination-polarization averaged version of the AK waveform. Additionally, we removed sources with initial eccentricity exceeding 0.9 and semi-latus smaller than 10 to avoid computational issues connected to the choice of the waveform.

We simulated the GW waveform of each EMRI assuming a 4 years mission and exploiting the innovative AAK waveform. We injected the GW signals in LISA using full TDI response. This technique combines the measurements along each LISA link in order to remove the laser frequency noise, which will be the dominant source of the noise.  Finally, we estimated the final GWB using an iterative algorithm which removes the resolvable sources from the background. 
We found that all the astrophysical models, except for one (the pessimistic M11), result in a detectable EMRI GWB, with SNRs accumulated in four years of observation ranging from $\approx20$ up to $>1000$, cf table \ref{tab3}. In the same timespan, LISA will allow the identification and characterization of several individual systems, up to $\approx 1000$ in the more optimistic case (again, cf table \ref{tab3}). We stress that since we excluded sources with initial semi-latus rectum smaller than 10, which can contribute significantly to the signal, the aforementioned numbers are likely  underestimated. The background will act mainly in the range $10^{-3}-10^{-2}\,\rm Hz$,  in particular, GWBs from optimistic models (such as M7 and M12) are loud enough to dominate over the LISA instrumental noise in this frequency range.

Our findings have several practical consequences for the forthcoming LISA mission. Specifically, EMRI GWB can significantly contribute to the LISA noise level around the sensitivity bucket, possibly jeopardizing the detectability of other interesting sources in the LISA window. In particular, the effect might be severe for two families of GWs sources that are of paramount importance for the mission, namely the merger of MBHB seeds occurring at high redshift and SOBH binaries. Because of their low mass and high redshift, seed binaries will populate the bottom of the LISA sensitivity bucket, overlapping with the EMRI GWB. SOBH binaries, instead, can be grouped into two categories: slow inspirals and multiband systems. The latter, which are perhaps the most interesting, are observed at $f>0.01$ Hz and should be unaffected. Conversely, slow inspirals are mostly found at $f<0.01$ Hz, where the GWB is more prominent and can greatly impact their SNR.

Finally, the EMRI GWB is not the only stochastic signal expected to fall into the LISA frequency band. Other sources of stochastic astrophysical foregrounds are anticipated to be GCBs and extragalactic NS and SOBH binaries. In our galaxy alone, it is estimated that there are roughly 250 million detached and 10 million interacting white dwarf binaries \cite{Edlund2005}. The sensitivity of LISA will allow the detection of thousands of these binaries as individual sources. The GWs from the other millions of binaries, especially below 1 mHz, will combine to form an unresolved background. Unlike EMRIs, these sources are anisotropically distributed in the sky and are concentrated in the disk of our Galaxy. As LISA orbits the sun, its orientation will change continuously, changing the antenna pattern of the detector and, consequently, the directions along which the detector will be most sensitive. The GW signals coming from the galactic plane, hosting the vast majority of WD binaries, will therefore fluctuate during a one-year cycle, making the identification of the WD noise easier. Conversely, extragalactic SOBH and NS binaries will also be isotropically distributed, and their resulting GWB will overlap significantly with the one produced by EMRIs. The detection and characterization of these astrophysical backgrounds is going to be challenging, and further studies and dedicated analysis pipeline developments are needed in order to assess the potential of LISA to identify and separate them from each other.

\section*{Acknowledgments}
A.S. acknowledges financial support provided under the European Union’s H2020 ERC Consolidator Grant ``Binary Massive Black Hole Astrophysics'' (B Massive, Grant Agreement: 818691). S.B.  acknowledges support from the French space
agency CNES in the framework of LISA.

\nocite{*}

\bibliographystyle{apsrev4-2}
\bibliography{apssamp}% Produces the bibliography via BibTeX.

\end{document}